\documentclass[aps,pra,10pt,letterpaper,twocolumn,superscriptaddress,showpacs,preprintnumbers]{revtex4-1}

\usepackage{etoolbox}
\usepackage{ulem}

\usepackage{color}
\usepackage{graphicx}
\usepackage{epstopdf}
\usepackage{bm}
\usepackage{dcolumn}
\usepackage{verbatim}
\usepackage{mathbbol}
\usepackage{dcolumn}
\usepackage{amsmath}
\usepackage{placeins}
\usepackage{mathrsfs}
\usepackage[colorlinks,linkcolor=red,citecolor=blue]{hyperref}

\DeclareTextFontCommand{\emph}{\it}

\newcommand{\APT}{{\scriptscriptstyle\mathrm{APT}}}

\newcommand{\IR}{{\scriptscriptstyle\mathrm{IR}}}
\newcommand{\SB}{{\scriptscriptstyle\mathrm{SB}}}
\newcommand{\HH}{{\scriptscriptstyle\mathrm{H}}}
\newcommand{\RABITT}{{\footnotesize\textsc{RABBIT }}}
\newcommand{\SBt}{{\footnotesize\textsc{SB}}}
\newcommand{\HHt}{{\footnotesize\textsc{H}}}

\definecolor{ReviseColorA}{rgb}{0.4,0,0.6}
\definecolor{ReviseColorB}{rgb}{0,0,0.8}
\newtoggle{ReviseToggle}

\definecolor{LightGrey}{rgb}{0.6,0.6,0.6}

\begin{document}

\title{Modulation of attosecond beating in resonant two-photon ionization}

\newcommand{\uam}{Departamento de Qu\'imica, M\'odulo 13, Universidad Aut\'onoma de Madrid, 28049 Madrid, Spain, EU}
\newcommand{\imdea}{Instituto Madrile\~no de Estudios Avanzados en Nanociencia (IMDEA-Nanociencia), Cantoblanco, 28049 Madrid, Spain, EU}
\author{\'Alvaro Jim\'enez-Gal\'an} \affiliation{\uam}
\author{Luca Argenti}\email{luca.argenti@uam.es} \affiliation{\uam}
\author{Fernando Mart\'in} \affiliation{\uam} \affiliation{\imdea}

\date{\today}
\toggletrue{ReviseToggle}

\begin{abstract}
We present a theoretical study of the photoelectron attosecond beating at the basis of \RABITT (Reconstruction of Attosecond Beating By Interference of Two-photon transitions) in the presence of autoionizing states. We show that, as a harmonic traverses a resonance, its sidebands exhibit a peaked phase shift as well as a modulation of the beating frequency itself. Furthermore, the beating between two resonant paths persists even when the pump and the probe pulses do not overlap, thus providing a sensitive non-holographic interferometric means to reconstruct coherent metastable wave packets. We characterize these phenomena quantitatively with a general finite-pulse analytical model that accounts for the effect of both intermediate and final resonances on two-photon processes, at a negligible computational cost. The model predictions are in excellent agreement with those of accurate \emph{ab initio} calculations for the helium atom in the region of the $N=2$ doubly excited states. 
\end{abstract}



\maketitle

Attosecond experiments~\cite{Krausz2009,Popmintchev2010,Sansone2011} can provide a time-resolved view~\cite{Leone2014} of the ultrafast electron dynamics that occurs in atoms and molecules (see, e.g., \cite{Sansone2010,Kelkensberg2011,Laurent2012}). A popular approach is \RABITT (reconstruction of attosecond beating by interference of two-photon absorption)~\cite{Paul2001b}. In this technique, a pump extreme-ultraviolet attosecond-pulse train (XUV-APT) is used in association with a compressed IR probe pulse to ionize the target atom or molecule and the photoelectron spectrum is recorded as a function of the pump-probe time delay, $\tau$.  
In the long-pulse limit, the APT only comprises odd multiples \HHt$_{2n+1}$ of the IR carrier frequency $\omega_\IR$. 
The transition amplitudes for the two alternative $\mathrm{A}+\gamma_{\HH_{2n\pm1}}\pm\gamma_{\IR}\rightarrow \mathrm{A}^++e^-$ paths, therefore, interfere giving rise to a sideband \SBt$_{2n}$ whose intensity oscillates with frequency $2\omega_\IR$, $\Delta I_{2n}\propto \cos(\varphi_{2n}+2\omega_\IR\tau)$, where the $\varphi_{2n}$ offset is given by the relative phase between the two consecutive \HHt$_{2n\pm 1}$ harmonics and the so-called relative atomic phase~\cite{Jimenez2013}. 
The \RABITT technique, therefore, can be used to reconstruct either the APT from the harmonic phases~\cite{Agostini2004}, if the atomic phases are known, or the atomic phases~\cite{Klunder2011a,Guenot2012,Dahlstrom2014}, if the APT shape is known.

An appealing perspective is to use attosecond technologies to investigate photoionization processes governed by electron correlation~\cite{Hattig2013,Lepine2014,Breidbach2005,Pabst2011a}. In particular, correlation is responsible for the Auger (or autoionization) decay of multiply-excited states, a resonant process that may require several tens of femtoseconds to complete. Consequently, as for bound states~\cite{Ranitovic2010,Swoboda2010,Ishikawa2011a,Shivaram2012a,Shivaram2013}, the presence of autoionizing states, either as intermediate or final states, can dramatically alter the atomic photoionization spectra~\cite{Lambropoulos1981a,Themelis2004,Tong2005,Zhao2005,Wickenhauser2006,Morishita2007,Sekikawa2008,Argenti2010,Chu2010b,Chu2011,Chu2012,Argenti2013,Chu2014}.
\RABITT has been used to investigate resonant processes in which the contribution of the direct-ionization amplitude to the resonant path is negligible, namely, when bound electronic states are directly excited by the XUV-APT (as in helium~\cite{Swoboda2010}) or when autoionizing vibronic states are populated by the XUV-APT without simultaneous excitation of the ionization continuum (as in the N$_2$ molecule~\cite{Caillat2011}). These \RABITT experiments \cite{Swoboda2010,Caillat2011} are compatible with a jump of $\pi$ in the sideband phase-shift as the energy of one of the adjacent harmonics traverses the resonant intermediate state. To our knowledge, \RABITT has never been used when both non-resonant continuum and resonant amplitudes contribute to the total ionization amplitude in similar amounts, a circumstance that is the rule more than the exception when atomic doubly-excited states are populated from the ground state. 
Furthermore, the lifetimes of autoionizing resonances are often comparable to or larger than the duration of the ultra-short pulses employed in common attosecond pump-probe schemes. As a consequence, even in perturbative conditions, a stationary regime is never achieved and the stationary models used to extract dynamical information from \RABITT are outright inapplicable. All the above, leads to obvious complications in the analysis of the \RABITT spectrum.
 
In this letter we theoretically analyze the effect of intermediate and final autoionizing states on the \RABITT photoelectron spectrum of He. To do so, we have solved the full dimensional time-dependent Schr\"odinger equation (TDSE) by using a nearly exact method \cite{Argenti2010,Lindroth2012,Jimenez2013} and interpreted these results in terms of an analytical time-resolved model, based on Fano's autoionization theory~\cite{Fano1961}, which, from a minimum set of parameters, is able to reproduce with great accuracy the \emph{ab initio} photoelectron spectrum for arbitrary pulses.  
We focus in particular on He doubly-excited states lying below the $N=2$ excitation threshold of the He$^+$ parent ion~\cite{Madden1963,Cooper1963a,Tanner2000}.
Our results show that, when intermediate autoionizing states are involved, the \RABITT photoelectron spectra does not follow the existing picture. First, as a consequence of the finite pulse duration, the frequency of the sideband oscillation is no longer $2\omega_\IR$ and displays a pronounced resonant modulation. Second, as a consequence of the interplay between the resonant and the continuum contribution to the resonant quantum path, the apparent local phase offset does not undergo a $\pi$ excursion anymore. When autoionization resonances appear in the final state, the photoelectron spectrum exhibits complex Fano asymmetry $q$ parameters~\cite{Bachau1986}, which are periodically inverted as a function of the pump-probe time delay. Finally, \RABITT oscillations persist even when the time delay is so large that the probe pulse does not overlap with the APT anymore. The local beating phase can then be used to reconstruct the coherent metastable wave packet created by the pump pulse.

\begin{figure}[t]
\begin{center}
\includegraphics[width=\linewidth]{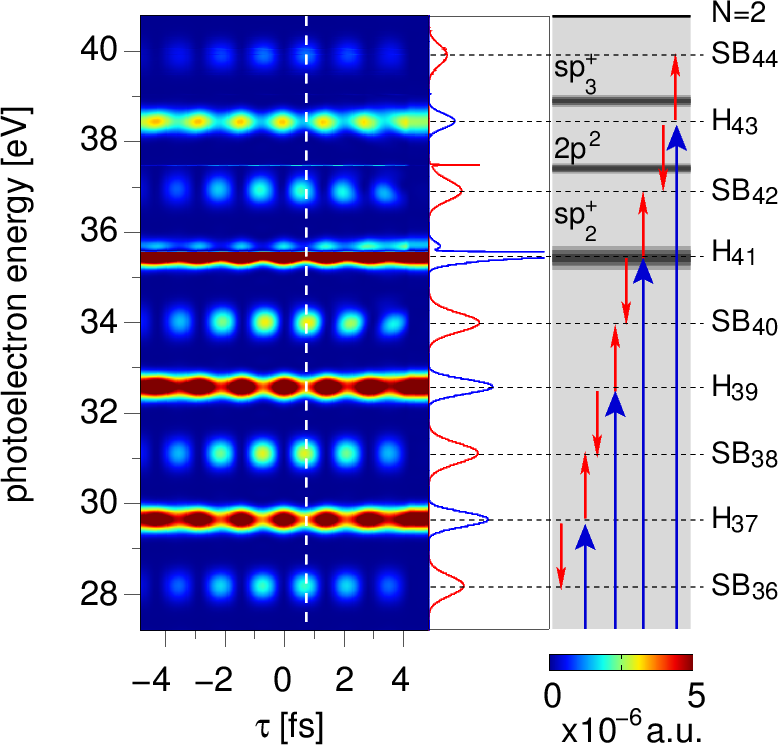}
\caption{\label{fig:EnergyScheme} (Color online)
Left panel: \emph{ab initio} photoelectron spectrum for the XUV-APT-pump weak-IR-probe ionization of helium, in the energy region of the N=2 DESs, as a function of the pump-probe time delay. The APT is centered at $\hbar\omega_\APT=57.21$~eV, $\hbar\omega_{\IR}=1.466$~eV, and both the pump and the probe have fwhm$\sim$6~fs. Central panel: sample signal at a fixed time delay (white dashed line). Right panel: outline of the most relevant states involved in the process. Starting from the $1s^2$ ground state, the atom absorbs one XUV photon from the APT and exchanges one or more IR photons with the probe pulse, giving rise to strong \HHt$_{2n+1}$ harmonic signals, in the {$^1$P$^o$} continuum, and to weak \SBt$_{2n}$ sideband signals, in the {$^1$S} and {$^1$D$^e$} continuum.}
\end{center}
\end{figure}

\begin{figure}[t]
\begin{center}
\includegraphics[width=\linewidth]{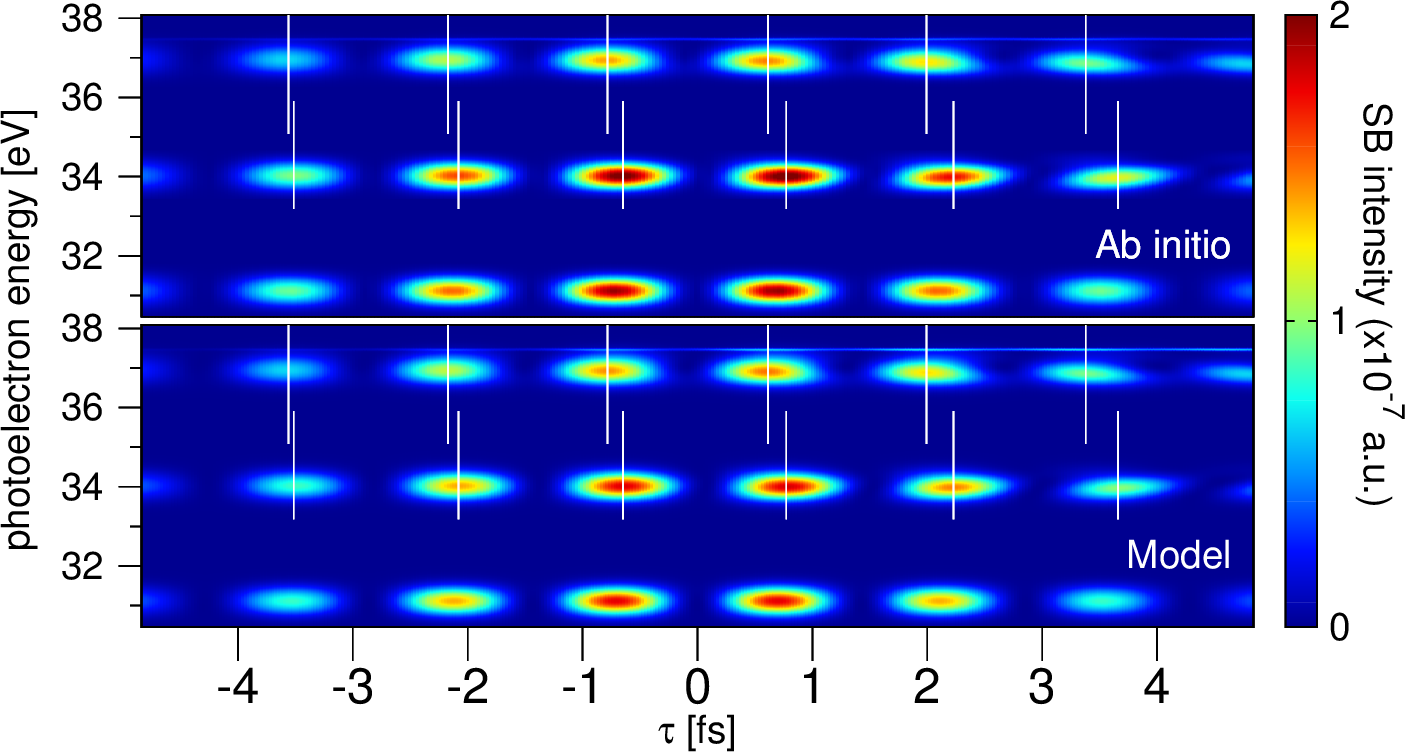}
\caption{\label{fig:SB_vs_td:OverlappingPulses} (Color online)
Calculated spectrum of the \SBt$_{38}$, \SBt$_{40}$ and \SBt$_{42}$ \RABITT sidebands, as a function of pump-probe time delay, for the fixed carrier energy $\hbar\omega_\IR=1.466$~eV. Top panel: \emph{ab initio} results. Bottom panel: results from the model (see text). The \SBt$_{42}$ beating \emph{frequency} is clearly higher than \SBt$_{40}$'s as a consequence of the intermediate $sp_2^+$ resonance being resonantly excited by the  \HHt$_{41}$ harmonic. }
\end{center}
\end{figure}

Fig.~\ref{fig:EnergyScheme} shows the \emph{ab-initio} results based on the solution of the TDSE for the helium atom ionized from the ground state with a train of extreme-ultra-violet (XUV) pulses in conjunction with a weak 800~nm infra-red (IR) pulse. 
Fig.~\ref{fig:SB_vs_td:OverlappingPulses} (top) just shows the \SBt$_{38-42}$ sidebands that arise when the $sp_2^+$ DES is resonantly excited by the \HHt$_{41}$ harmonic ($\hbar\omega_\IR=1.466$~eV). One can clearly see an apparent local phase shift of the sidebands that increases linearly with the time delay. Stated otherwise, the resonance alters the \RABITT beating \emph{frequency} itself.

To understand this behavior, we have developed a model in which the finite-pulse formulation of the second-order transition amplitude is used, 
\begin{equation}\label{eq:2ndOrderTDPT} 
\mathcal{A}^{(2)}_{f\gets i}(\omega_{fi})=-i\int_{-\infty}^\infty d\omega \tilde{A}(\omega_{fi}-\omega)\tilde{A}(\omega)\mathcal{M}_{fi}(\omega),
\end{equation}
where $\mathcal{M}_{fi}(\omega)=\alpha^2\langle \psi_f|P_z G^+_0(\omega)P_z|\psi_i\rangle$ is the two-photon transition kernel in velocity gauge, $\alpha$ is the fine-structure constant, $G^+_0(E)=(E-H+i0^+)^{-1}$ is the retarded resolvent of the field-free  hamiltonian, and $\tilde{A}(\omega)=(2\pi)^{-1/2}\int dt A(t) \exp(i\omega t)$ is the Fourier transform of the vector potential amplitude along the polarization axis (unless otherwise stated, atomic units are assumed).
The intermediate and final continuum states can be expressed as solutions of the single-channel Fano scattering problem~\cite{Fano1961}
\begin{equation}\label{eq:FanoExpression}
\begin{split}
|\psi_{E}^+\rangle 
=&|E\rangle + \left(\int \hspace{-2pt}d\varepsilon|\varepsilon\rangle\frac{V_{\varepsilon a}}{E-\varepsilon+i0^+}+|a\rangle\right)\frac{V_{aE}}{E-\tilde{E}_a}
\end{split}
\end{equation}
where $|E\rangle$ is the unperturbed continuum, and $|a\rangle$ is the localized part of an isolated autoionizing state with complex energy $\tilde{E}_a=E_a+\Delta_a-i\Gamma_a/2$. The replacement of (\ref{eq:FanoExpression}) in (\ref{eq:2ndOrderTDPT}) gives rise to a profusion of terms, all of which, however, can be expressed analytically in terms of a minimal set of transition-strength parameters, if we assume external Gaussian pulses and make a few additional reasonable approximations (see Supplementary Material). 
In the case of a single intermediate resonance $|a\rangle$, for example, the total transition amplitude $\mathcal{A}_{f\gets i}^{(2)}$ is given by the sum of two dominant contributions, $\mathcal{A}^{\pm}$, which correspond to the absorption of an XUV photon from the \HHt$_{2n\pm1}$ harmonics followed by the coherent emission or absorption of an IR photon, and have the following simplified expression
\begin{equation}\label{eq:Amplitudes}
\mathcal{A}^{\pm}= \mathcal{F}(\tau)\,e^{\mp i\omega_\IR\tau} 
\left[w(z_f^\pm)
+\frac{\beta_{Ea}-1}{\epsilon_{fa}}\,(q_a'-i ) \,w(z_a^\pm)\right],\nonumber
\end{equation}
where $\mathcal{F}(\tau)$ is a factor common to both $\mathcal{A}^{\pm}$ amplitudes and $w(z)$ is the Faddeeva special function~(see \S 7.1.3 in \cite{AbramowitzStegun}). The complex parameters $z_{a}^\pm$ and $z_f^\pm$ depend linearly on time delay and express the dephasing, accumulated across the pump-probe time overlap, of the effective harmonics \HHt$_{2n\pm1}$ with respect to the excitation energy of the autoionizing state $|a\rangle$ and the continuum $|f\rangle$, respectively (for the definition of the terms, see Supplementary Material~\cite{SupplementaryNotes}).
The energy-resolved intensity of the sideband is given by the square module of the total amplitude
\begin{equation}
I_{\SB}(\tau)=\left|\mathcal{A}^+(\tau)\right|^2+\left|\mathcal{A}^-(\tau)\right|^2+2\Re e\left[\mathcal{A}^{+*}(\tau)\mathcal{A}^-(\tau)\right].
\end{equation}
The local phase $\varphi(\omega_\IR,\tau)$ of the sideband fast modulation is nothing more than the argument of the complex interference term $\mathcal{A}^{+*}(\tau)\mathcal{A}^-(\tau)$ and comprises the $\RABITT$ phase $2\omega_\IR\tau$ as well as an additional dephasing, $\delta\varphi(\omega_\IR,\tau)$, which depends on both the frequency of the laser and on the time delay. For the cases we examined, the phase deviation is well approximated by a linear interpolation, $\delta\varphi(\omega_\IR,\tau)\simeq \delta\varphi_0(\omega_\IR)+\delta\omega(\omega_\IR)\tau$. The local phase shift, therefore, is affected by the apparent phase shift at $\tau=0$ $\varphi_0(\omega_\IR)$ as well as by the modulation  $\delta\omega(\omega_\IR)$ of the beating frequency. From the complete analytical expression for the $\mathcal{A}^\pm$ amplitudes, it is a simple task to derive the ultimate observable $\delta\varphi(\omega_\IR,\tau)$ as a function of the initial minimal set of atomic parameters.
Conversely, all these parameters can be determined by comparing the model prediction with few selected experiments~\cite{SupplementaryNotes}. We applied this latter procedure to helium by using our nearly exact \emph{ab initio} solutions of the TDSE ~\cite{Argenti2010,Lindroth2012,Jimenez2013} as ``numerical experiments''.

Fig.~\ref{fig:SB_vs_td:OverlappingPulses} (bottom) shows the results obtained with our model. As can be seen, the agreement with the \emph{ab initio} results is excellent. In particular, the change in the beating \emph{frequency} of the sidebands is very well reproduced.
We notice that the finite duration of the pulses induces a finite frequency detuning $\delta\omega(\omega_\IR)$ even in absence of intermediate resonances. For example, for $800$~nm $6$~fs IR pulses, the frequency detuning predicted by the linearized analytical formula is quite large, $\delta\omega=-0.0012$, in good agreement with both the value obtained with the full analytical model, $\delta\omega^{FT}=-0.0014$, and with that resulting from the \emph{ab initio} calculation $\delta\omega_{ai}=-0.0014$, both computed by Fourier analyzing the energy-integrated sideband signal. 
When \HHt$_{41}$ is resonant with the $sp_2^+$ state we obtain $\delta \omega\simeq-0.003$~a.u., in good agreement with the full-model value $\delta\omega^{FT}=-0.0027$~a.u.  Taking into account the background non-resonant frequency detuning, this means a frequency modulation as large as $0.0013$~a.u., i.e., $1.2\%$ of $2\omega_\IR$, which corresponds to a phase-shift gradient of almost $10^\circ$ per laser cycle, or to a change in the \RABITT period of 17~as.

\begin{figure}[t]
\begin{center}
\includegraphics[width=\linewidth]{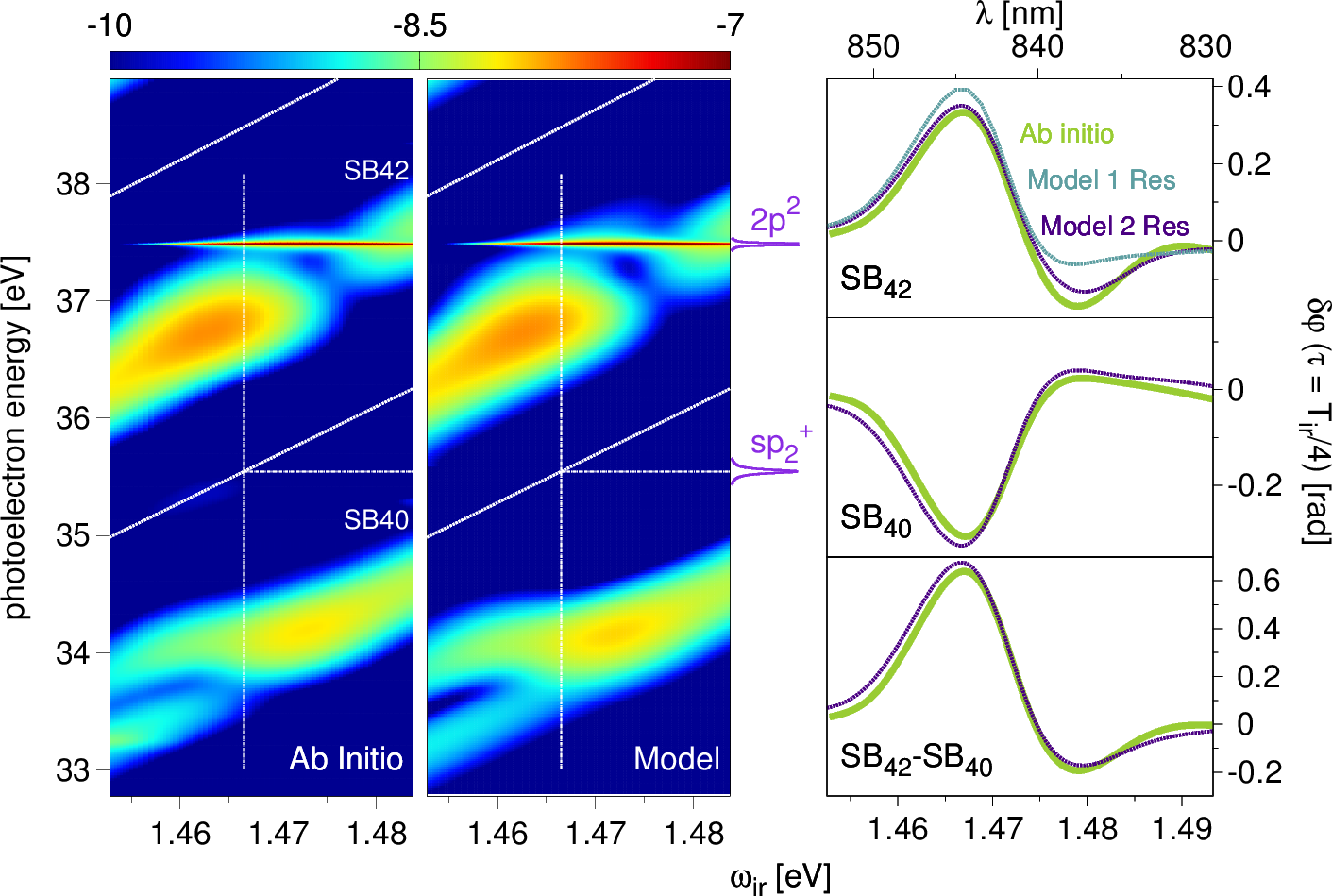}
\caption{\label{fig:SB_vs_Omega:td0} (Color online) On the left,
sidebands \SBt$_{40}$ and \SBt$_{42}$  vs. $\omega_\IR$, for $\tau=0$, computed both \emph{ab initio} and with the model. On the right, comparison of the energy-integrated signals (magenta dotted line: model prediction; green solid line: \emph{ab initio} calculation). The $sp_2^+$ resonance gives rise to a large peak in the apparent dephasing between the two sidebands. 
}
\end{center}
\end{figure}
In the special case of a dark non-resonant continuum ($q_a'\to\infty$), the model reproduces the characteristic $\pi$ jump discussed in~\cite{Swoboda2010,Caillat2011}.  In general, however, the concurrence of a resonant and a non-resonant amplitude results in a qualitatively different profile. Fig.~\ref{fig:SB_vs_Omega:td0} shows the energy-resolved \SBt$_{40}$ and \SBt$_{42}$ sidebands, computed either \emph{ab initio} or with the model, as a function of the IR photon energy for a fixed time delay, $\tau=0$, close to the minimum of the beating. In this case, the model included the two bright $sp_{2/3}^+$ intermediate $^1$P$^o$ resonances as well as the final {$^1$S} $2p^2$ resonance. The first two plots, which are in very good agreement, illustrate well two aspects of the effects of resonances in \RABITT experiment. First, while the upper sideband displays a maximum at negative and a minimum at positive detuning of the \HHt$_{41}$ resonant harmonic energy with respect to the excitation energy of the $sp_2^+$ state from the ground state, the opposite is true for the lower sideband. 
 Second, due to the strong $sp_2^+-2p^2$ coupling, the $2p^2$ state already starts populating when the $sp_2^+$ state is resonant with the \HHt$_{41}$ harmonic ($\hbar\omega_\IR=1.466$~eV), i.e., well before the $2p^2$ state is resonant with the \SBt$_{42}$ sideband ($\hbar\omega_\IR=1.478$~eV).
 
In \SBt$_{42}$, we can also recognize the clear signature of the upper $sp_3^+$ resonance, which becomes resonant at larger frequencies thus inducing a more pronounced sigmoidal profile. Indeed, inclusion of the $sp_3^+$ state brings the model in much better agreement with the simulation. The profile of the apparent phase shift is clearly very different from the $\pi$ jump mentioned earlier; this aspect remains true even in the stationary limit ($\sigma_t\to\infty$), when the frequency modulation $\delta\omega(\omega_\IR)$ vanishes.  
In any case, in these conditions, adjacent sidebands have an apparent dephasing as large as 0.8~rad which should be comfortably observable with existing technology~\cite{SalieresPrivCom}. 

\begin{figure}[t]
\begin{center}
\includegraphics[width=\linewidth]{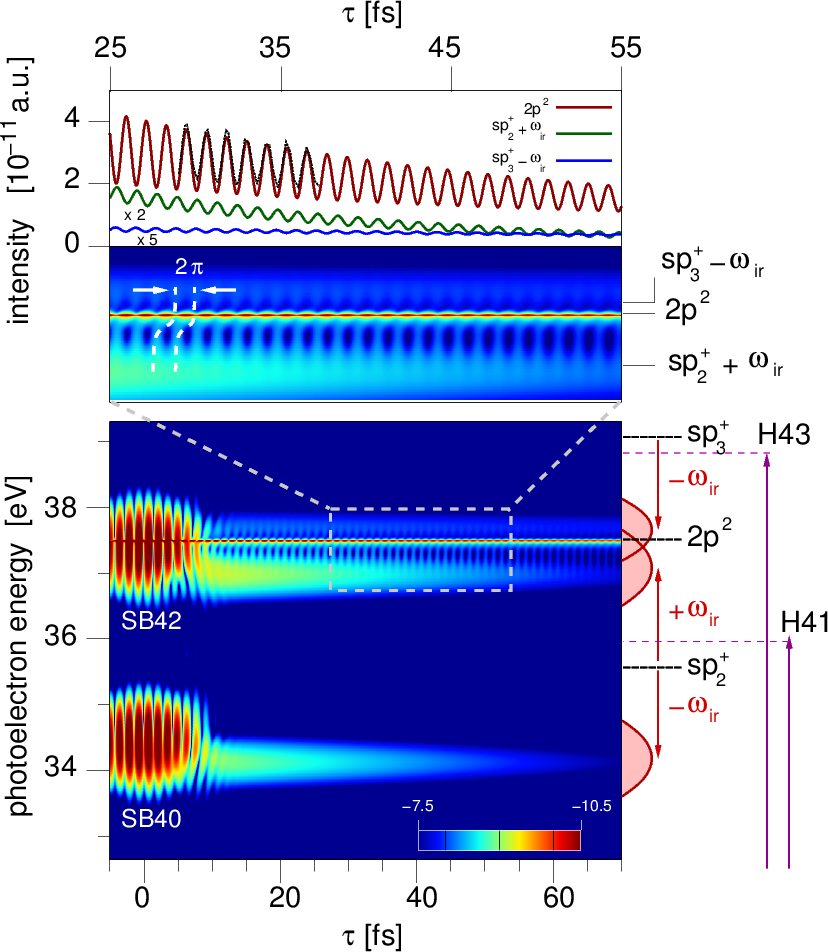}
\caption{\label{fig:SB_vs_td:SeparatePulses} (Color online) Bottom panel: spectrum of the \SBt$_{40,42}$ sidebands vs. $\tau$ computed with the model, for $\hbar\omega_\IR=1.475$~eV. The harmonics \HHt$_{41}$ and \HHt$_{43}$ (not shown) are detuned from the $sp_{2}^+$ and $sp_3^+$ {$^1$P$^o$} DES by $\delta_{sp_2^+}=0.37$~eV and $\delta_{sp_3^+}=-0.19$~eV, respectively. When the APT and the IR pulse overlap ($|\tau|<5$~fs), the sidebands are dominated by the non-resonant signal and centered at $E=2n\omega_\IR-IP$. Between 5~fs and 10~fs, the non-resonant contributions disappear, while the sidebands narrow and shift to symmetric positions around the two resonances. For larger time delays, the \SBt$_{42}$ sideband displays the characteristic interference fringes of the $sp_2^+-sp_3^+$ beating, with a lifetime intermediate between those of the two resonances. The final $2p^2$ resonant signal, which dominates the spectrum and reproduces the \emph{ab initio} prediction (black thin solid line) with high accuracy, oscillates in anti-phase with respect to the background (inversion of the Fano profile).}
\end{center}
\end{figure}

Fig.~\ref{fig:SB_vs_td:SeparatePulses} shows the \SBt$_{40,42}$ sidebands for large time delays, obtained by using a slightly different laser frequency ($\omega_{IR}=1.475$~eV). In contrast with non-resonant two-photon transitions, the sideband signals persist even when pump and probe do not overlap, with the exponentially decaying sidebands being located symmetrically with respect to the intermediate $sp_2^+$ and $sp_3^+$ autoionizing states.  
These states, of {$^1$P$^o$} symmetry, are populated by the harmonics \HHt$_{41}$ and \HHt$_{43}$, thus giving rise in the intermediate energy region to a signal beating at the complex frequency $\tilde{E}_{sp_3^+}-\tilde{E}_{sp2^+}^*$ where the sidebands from the two resonances overlap.
The strongest contribution to the sideband comes from the transition to the $2p^2$ state, which is permitted even at the level of the independent-particle approximation. The (complex) $q$  parameters for the excitation of the $2p^2$ state from either the $sp_2^+$ or the $sp_3^+$ states differ, thus giving rise to a beating of the background continuum that is out of phase with respect to that of the final resonance.
In the case of two-photon transitions, a final resonance appears in the spectrum as a Fano profile with a complex $q$. Here the Fano profile is inverted periodically as a function of the pump-probe time delay. This effect is similar to the one observed in attosecond transient absorption spectroscopy~\cite{Ott2012,Argenti2012f,Ott2013}. Indeed, the oscillation is the signature of the coherence between the two intermediate resonances; thus, its phase can in principle provide us a photoelectron interferometric way to reconstruct the autoionizing wave packet itself, alternative to those based on the holographic principle~\cite{Mauritsson2010}.

In conclusion, we have solved the TDSE and developed an analytical model for the two-photon ionization of atoms with finite pulses, in the presence of autoionizing states. The model shows that both the contribution of intermediate continuum states and the finite duration of the light-pulses must be taken into account to achieve qualitatively correct interpretations of resonant attosecond pump-probe experiments. 
In particular, we have demonstrated that: i) intermediate resonances manifest themselves in \RABITT experiments with variations in the sideband apparent phase shift and the beating frequency, as a function of the fundamental carrier frequency; ii) resonances in the final states appear in the photoelectron spectrum as Fano-like profiles whose $q$ parameters are complex, strongly modulated with respect to the pump-probe time delay ($q$ inversion), and out of phase with respect to the background signal. From a simplified version of the model, we have derived a short expression for the sideband beating, based on the Faddeeva special function. For the realistic cases we have examined, the variation of the apparent sideband phase shift is larger than 0.5~rad and should therefore be easily detectable with current instrumental resolution.
Despite its essential and analytical form, the model is able to provide results in quantitative agreement with accurate \emph{ab initio} solutions of the TDSE, thus permitting us to explore a vast range of pulse parameters at a negligible computational cost or, conversely, to extract radiative-transition strengths between multiply excited states, which are hard to obtain otherwise. 
Finally, long-lived resonances excited by consecutive harmonics give rise to sideband beatings that persist even when the pump and probe pulses do not overlap and from which the coherent metastable wave packet can be reconstructed.

We thank Richard Ta\"ieb, Alfred Maquet and Jeremie Caillat for useful discussions and their kind hospitality. We acknowledge support from the European Research Council under the European Union's Seventh Framework Programme (FP7/2007-2013)/ERC grant agreement 290853, the MICINN project FIS2010-15127, the ERA-Chemistry Project PIM2010EEC-00751, the European grant MC-ITN CORINF and the European COST Action XLIC CM1204.

\FloatBarrier


\end{document}